\documentclass[aps,prb,showpacs,twocolumn]{revtex4-1}

\usepackage{graphicx}
\usepackage{dcolumn}
\usepackage{color}
\usepackage{amsmath,amssymb,bm, array,tabularx,booktabs,multirow}
\usepackage{longtable}
\usepackage{braket}
\usepackage{here}

\usepackage[normalem]{ulem}

\begin{document}

\setlength{\extrarowheight}{4pt}

\title{Piezo-optic effect of high-harmonic generation in semiconductors}
\author{Tomohiro Tamaya}
\email{tamaya@issp.u-tokyo.ac.jp}
\author{Takeo Kato}
\affiliation{
${}$Institute for Solid State Physics, University of Tokyo, Kashiwa, 277-8581, Japan}

\date{\today}

\begin{abstract}
We theoretically investigate the piezo-optic effect of high-harmonic generation (HHG) in shear-strained semiconductors. 
By focusing on a typical semiconductor, GaAs, we show that there is optical activity, meaning different responses to right-handed and left-handed elliptically polarized electric fields. 
We also show that this optical activity is more pronounced for higher harmonics whose perturbative order exceeds the band-gap energy. 
These findings point to a useful pathway for strain engineering of nonlinear optics to control the reciprocity of HHG.
\end{abstract}

\maketitle

\section{Introduction}
High-harmonic generation (HHG) is one of the most fundamental topics in nonlinear optics\cite{Boyd2008,Yariv1984,Shen1984}. 
Experimental progress from the perturbative to the nonperturbative regime in gaseous media has paved the way for developing novel optical devices for, e.g., generating short-wavelength attosecond pulses \cite{Corkum1993,Protopapas1997,Brabec2000,Agostini2004,Corkum2007,Krausz2009}. 
Moreover, HHG in the nonperturbative regime has been experimentally observed in solids\cite{Ghimire2011,Schubert2014,Luu2015,Hohenleutner2015,Vampa2015,Ghimire2019}, and ensuing studies have opened up a new field in condensed-matter science~\cite{Ndabashimiye2016,Lanin2017,Liu2017,You2017,Yoshikawa2017,Lanin2017,Kim2017,Jiang2018,Langer2018,VampaOE2018,Silva2018,Hirori2019,Bing2020}. 
In contrast to gaseous media, HHG in solids has various inherent properties that are rooted in the crystallinity of the medium and may provide means of developing new optical devices that use HHG. 
Thus, it is important to study the characteristics of HHG in various materials and to devise a control method that can provide a possible route to novel optical technology.

The most important aspects that determine the properties of HHG are the band structures of the materials and the corresponding Bloch wavefunctions. 
The Hamiltonian of the light-matter interaction is principally made up of these elements and HHG is expected to yield unusual new features by appropriately choosing those materials. 
A recent experimental study reported that HHG in monolayer $\rm{MoS}_{2}$ {was} polarized perpendicular to the linearly polarized pump field\cite{You2017}, an effect that was mainly explained in terms of the anomalous transverse intraband current arising from the material's Berry curvature. 
Thus, the properties of {the} Bloch wavefunctions in materials directly affect the characteristics of HHG, and {thus}, exploring methods of controlling these wavefunctions are crucial for applications of HHG.

\begin{figure}[tb]
\begin{center}
\includegraphics[width=8cm]{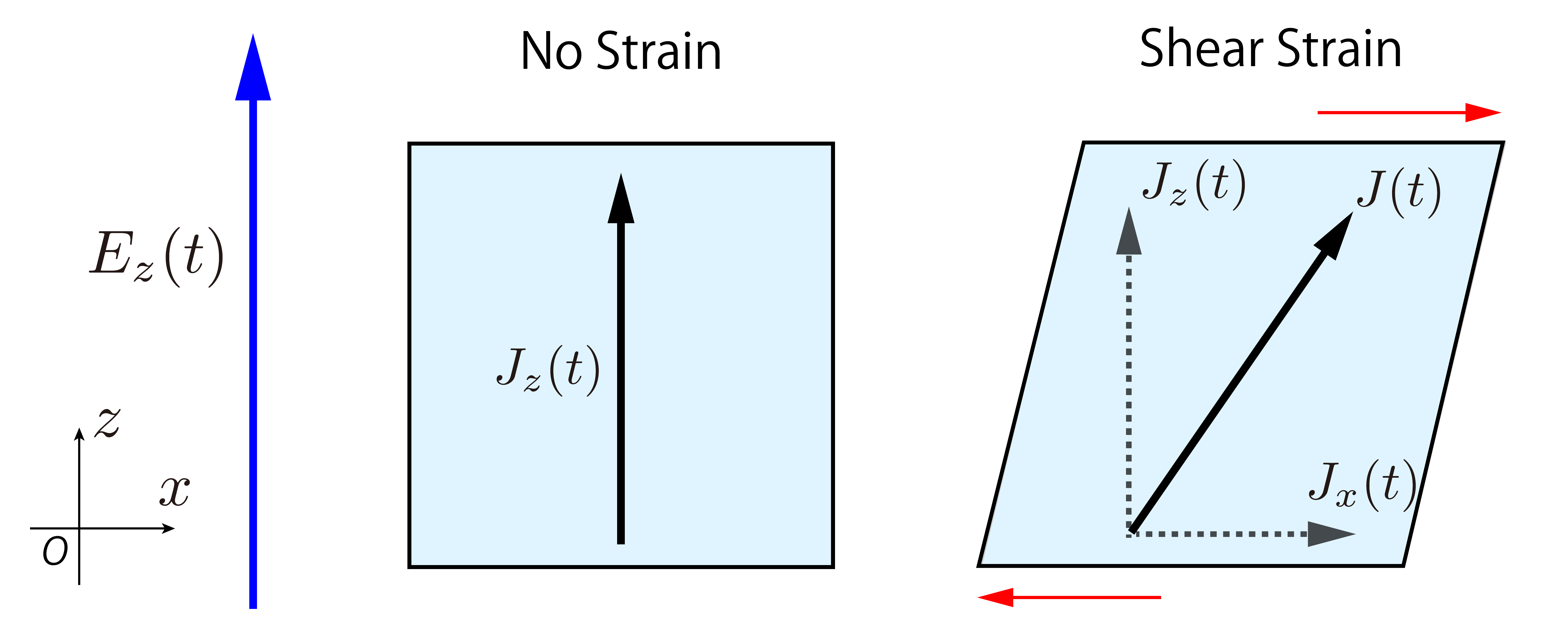}
\caption{(Color online) Schematic diagram of the current $\bm{J}(t)$ generated by a linearly polarized electric field. The left figure indicates the generated current in the absence of strain, while the right figure indicates the generated current when an external shear strain is applied. In the absence of strain, current is generated only in the $z$-direction parallel to the electric field of the incident light. When shear strain is applied to the system, the direction of the current is distorted and $\bm{J}(t)$ has both $z$- and $x$-components.
\label{fig:config1}}
\end{center}
\end{figure}

One possible way to control the Bloch wavefunctions in materials is strain engineering\cite{smith1954piezoresistance,llordes2012nanoscale,lu2016enhanced,liu2017pressure,chen2018large,zhao2017strained,zhu2019strain,steele2019thermal,peng2020strain,conley2013bandgap,he2013experimental,frank2011raman,shi2013quasiparticle,castellanos2015mechanics,bertolazzi2011stretching,castellanos2012elastic,bao2017low,chen2020strain,xiong2020strain,shi2019strain,wang2015}. 
Mechanical deformation of a material modifies the Bloch wavefunction{s} by distorting the crystal structure, and it can be used to control various physical properties such as transport and optical response. 
{W}e expect, for example, that shear strain will rotate the direction of the generated current (see Fig.~\ref{fig:config1}). 
This rotation of the current direction indicates left-right symmetry breaking in materials leading to emergent optical activity of HHG, i.e., different responses to right-handed and left-handed elliptically polarized electric fields\cite{barron2009molecular,mason1982molecular}. 
In particular, the piezo-optic effect of HHG referred to here could be used for applications such as spatially resolved distortion measurement and mechanical control of HHG, which are considered impossible for gaseous media\cite{wang2020strain,shao2019strain}.

In this paper, we theoretically investigate the characteristics of HHG in shear-strained semiconductors. 
We construct a theoretical framework based on the Luttinger-Kohn-Bir-Pikus Hamiltonian\cite{Luttinger1955,Ahn1995,Dargys2002,Pryor1998,Tomic2006,Chuang1991,Pfeffer1996,sytnyk2018,BirPikus1974,luque2015comparing}, which provide{s} us with a general platform to treat various semiconductors under external shear strain. 
By performing a dynamical simulation on GaAs{,} a typical III-V semiconductor, we find that external shear strain indeed causes different responses {to} right-handed and left-handed elliptically polarized electric fields. 
We also find that th{is} non-reciprocity is more pronounced for higher harmonics whose perturbative order exceeds the band-gap energy.

The organization of this paper is as follows. 
In Sec.~\ref{Formulation}, we introduce the theoretical framework for HHG using the Luttinger-Kohn-Bir-Pikus Hamiltonian. 
In Sec.~\ref{RESULTS AND DISCUSSIONS}, we show numerical results for HHG emitted from shear-strained GaAs. 
We also discuss the optical activity of HHG in detail {by examining} the different optical responses to right- and left-handed elliptically polarized light. 
Sec.~\ref{Conclusion} summarizes the conclusions of this study.
{In the appendix, we discuss the ellipticity of emitted harmonics.}

\section{Formulation}
\label{Formulation}
The general formulation introduced here for HHG is applicable to various semiconductors with a direct band gap at the $\Gamma$ point. 
First, we describe the eight-band Luttinger-Kohn model in Sec.~\ref{sec:KohnLuttingerModel}; then, we extend it to the strained case, called the Pikus-Bir Hamiltonian model, in Sec.~\ref{sec:PikusBirHamiltonian}. 
Next, we derive the light-matter interaction in terms of the Luttinger-Kohn-Pikus-Bir model in Sec.~\ref{sec:LightMatterInteraction}.
Finally, we describe the time-dependent Schr\"odinger equation for electrons and define the polarization currents in Sec.~\ref{sec:dynamics}.

\subsection{Luttinger-Kohn model}
\label{sec:KohnLuttingerModel}
Let us consider a general microscopic Hamiltonian,
\begin{align}
H =\frac{\bm{p}^2}{2m_0} +\sum_{i}\left[ V_i(\bm{x}) + \frac{\hbar}{4m_0^2 c^2}
(\nabla V_i \times {\bm p})\cdot {\bm \sigma}
\right],
\end{align} 
where $m_{0}$ is the electron mass, $\bm{p}$ is the momentum of the electron, and $V_i({\bm x})=V(\bm{x}-\bm{R}_{i})$ is the periodic core potential of atoms located at $\bm{R}_{i}$. 
The second term in brackets expresses the spin-orbit coupling, where ${\bm \sigma}$ is the spin angular momentum. By performing a band calculation, the Hamiltonian can be diagonalized as
\begin{align}
H |\Psi_{n{\bm k}}\rangle = E_{n{\bm k}} |\Psi_{n{\bm k}}\rangle,
\end{align}
where $n$ is the band index, ${\bm k}$ is the Bloch wavenumber, $E_{n{\bm k}}$ is the energy dispersion of the $n$th band, and $|\Psi_{n{\bm k}}\rangle$ is the Bloch wavefunction. 
Here, we will focus on the bands near the band edge at the $\Gamma$ point (${\bm k}=0$) and restrict them to eight bands composed of one conduction band ($n=1$) and three valence bands, i.e., a heavy-hole band ($n=2$), a light-hole band ($n=3$), and a split-off band ($n=4$), and their time-reversal counterparts ($n=5,6,7,8$).

We apply conventional ${\bm{k}} \cdot {\bm{p}}$ perturbation theory\cite{harrison2012electronic,kittel1963quantum,peter2010fundamentals,voon2009kp} around the $\Gamma$ point using these eight bands. 
We rewrite the Bloch wavefunction as $\ket{\Psi_{n\bm{k}}}= e^{i {\bm{k}} \cdot {\bm{x}}} \ket{u_{n{\bm{k}}}}$. 
The eigenvalue equation is rewritten as $\tilde{H} \ket{u_{n{\bm{k}}}} = E_{n{\bm k}} \ket{u_{n{\bm{k}}}}$, where $\tilde{H}$ is an effective Hamiltonian defined as
\begin{align}
\tilde{H} &\equiv e^{-i {\bm{k}} \cdot {\bm{x}}} H e^{i {\bm{k}} \cdot {\bm{x}}} 
\equiv \tilde{H}_0 + \tilde{V}, \\ 
\tilde{H}_0 &= \frac{{\bm {p}}^2}{2 m_{0}} +\sum_{i}\left[ V_i(\bm{x}) + \frac{\hbar}{4m_0^2 c^2}
(\nabla V_i \times {\bm p})\cdot {\bm \sigma}
\right], \\
\tilde{V} &=\frac{\hbar}{m_{0}}{\bm{k}} \cdot {\bm{p}} + \frac{\hbar^2 \bm{k}^2}{2 m_{0}}. 
\end{align}
The unperturbed Haimiltonian $\tilde{H}_0$ is diagonalized by the wavefunction at the $\Gamma$ point, $\ket{u_n} \equiv \ket{u_{n{\bm k}=0}}$. 
Following conventional $\bm{k} \cdot \bm{p}$ perturbation theory, we incorporate the ${\bm k}$-dependence of the eigen wavefunctions by second-order perturbation with respect to $\tilde{V}$, taking the effect of outside bands other than the target bands into account. 
The resultant effective Hamiltonian is \cite{Tomic2006}
\begin{align}
\label{HH}
\braket{u_n| \tilde{H}^{\rm{eff}}_{0} |u_{n'}} = \left( \begin{array}{cc} 
H^{\bm k}_{uu} & H^{\bm k}_{ul} \\ H^{\bm k}_{lu} & H^{\bm k}_{ll}
\end{array} \right),
\end{align}
where $H^{\bm{k}}_{uu}$, $H^{\bm{k}}_{ul}$, $H^{\bm{k}}_{lu}$, and $H^{\bm{k}}_{ll}$ are 4 $\times$ 4 submatrices. The submatrix $H^{\bm{k}}_{uu}$ has the form,
\begin{eqnarray} 
H^{\bm k}_{uu} = \left( \begin{array}{cccc} 
E_{CB} & -\sqrt{3}T & \sqrt{2}U & -U \\ -\sqrt{3}T^{*} & E_{HH} & \sqrt{2}S & -S \\ \sqrt{2}U & \sqrt{2}S^{*} & E_{LH} & -\sqrt{2}Q \\ -U & -S^{*} & -\sqrt{2}Q & E_{SO}\\
\end{array} \right), 
\label{Hamiltonian1}
\end{eqnarray}
while the submatrix $H_{ll}$ is defined as $H_{ll}=H_{uu}^{*}$. The submatrices, $H_{ul}$ and $H_{lu}$, are expressed as
\begin{eqnarray}
H^{\bm{k}}_{ul} = \left( \begin{array}{cccc} 
0 & 0 & -T^{*} & -\sqrt{2}T^{*} \\ 0 & 0 & -R & -\sqrt{2}R \\ T^{*} & R & 0 & \sqrt{3}S \\ \sqrt{2}T^{*} & \sqrt{2}R & -\sqrt{3}S & 0\\
\end{array} \right),
\label{eq:Hamiltonian2}
\end{eqnarray}
and {$H_{lu}=H_{ul}^{\dagger}$}. The diagonal elements of $H_{uu}$ and $H_{ll}$ are defined as
\begin{align}
E_{CB}& = E_{g} + O, \\
E_{HH}& = - (P + Q), \\
E_{LH}& = - (P - Q), \\
E_{SO}& = - (P + \Delta_{SO}). \label{eq:Hamiltonian3}
\end{align}
The subscripts CB, HH, LH, and SO stand for conduction, heavy-hole, light-hole and split-off bands, respectively, and $E_{g}$ and $\Delta_{SO}$ are the band-gap energy and the split-off energy {due to the spin-orbit interaction}, where $E_{g}=1.42$ eV and $\Delta_{SO}=0.34$ eV in GaAs.

In the absence of external strain, the matrix elements are 
\begin{align}
O &= \frac{\hbar^2}{2 m_{0}}\gamma_{C}\left(k_{x}^2 + {k}_{y}^2 + {k}_{z}^2 \right), \\
P & = \frac{\hbar^2}{2 m_{0}}\gamma_{1}\left(k_{x}^2 + {k}_{y}^2 + {k}_{z}^2 \right), \\
Q & = \frac{\hbar^2}{2 m_{0}}\gamma_{2}\left(k_{x}^2 + k_{y}^2 - 2 k_{z}^2 \right), \\
R & = \frac{\hbar^2}{2 m_{0}}\sqrt{3}\left[\gamma_{2}(k_{x}^2 - k_{y}^2) - 2 i \gamma_{3} k_{x} k_{y} \right], \\
S & = \frac{\hbar^2}{2 m_{0}}\sqrt{6}\gamma_{3}\left(k_{x} - i k_{y}\right)k_{z}, \\
T & = \frac{1}{\sqrt{6}}P_{0}\left(k_{x} + i k_{y}\right), \\
U & = \frac{1}{\sqrt{3}}P_{0} k_{z}.
\label{eq:Hamiltonian4}
\end{align}
Here, $k_{x}$, $k_{y}$, and $k_{z}$ denote components of the Bloch wavevector along the [100], [010], and [001] crystallographic directions, respectively, and $\gamma_{0}$, $\gamma_{1}$, $\gamma_{2}$, and $\gamma_{3}$ are the Luttinger parameters. 
We set the Luttinger parameters of GaAs to be $\gamma_{C}=0.5$, $\gamma_{1}=2.7$, $\gamma_{2}=-0.1$, and $\gamma_{3}=0.7$ following Ref.~\onlinecite{sytnyk2018}. 
The dipole matrix element (the Kane matrix element) is defined as
\begin{equation}
P_{0} = -i\left(\frac{\hbar}{m_{0}}\right)\braket{s;\sigma|p_{\lambda}|\lambda;\sigma},
\end{equation}
where $\lambda = x, y, z$. {The value of $P_0$ can be absorbed into the definition of the Rabi frequency introduced later.}

\subsection{Strain-induced effect in semiconductors}
\label{sec:PikusBirHamiltonian}
Strain in the crystal is expressed by displacement of the lattice vectors from {those} of the unstrained crystal, {${\bm x}_i$} ($i=x,y,z$):
\begin{align}
\delta x^{j}_{i} = \sum_{i} \delta_{ij} x^{j}_{i},
\end{align}
where {$x^j_i$ is the $j$-th component of ${\bm x}_i$ ($j=x,y,z$), and} $\delta_{ij}$ ($i,j=x,y,z$) denote components of the strain tensor. 
The effect of the strain can be incorporated into the ${\bm k} \cdot {\bm p}$ band structure calculations by adding an extra perturbation term to the unstrained potential\cite{BirPikus1974}. 
Thus, the change in the matrix elements in the presence of the strain is obtained as $O \rightarrow O + \delta O$, $P\rightarrow P+\delta P$, and so on, where\cite{Luttinger1955,Ahn1995,Dargys2002,Pryor1998,Tomic2006,Chuang1991,Pfeffer1996,sytnyk2018,BirPikus1974,luque2015comparing}
\begin{align}
\delta O & = + a_c \left(\delta_{xx} + \delta_{yy} + \delta_{zz} \right), \\
\delta P & = - a_v (\delta_{xx} + \delta_{yy} + \delta_{zz}), \\
\delta Q & = -\frac{b_v}{2}\left(\delta_{xx} + \delta_{yy} + \delta_{zz} \right), \\
\delta R & = -\frac{\sqrt{3}}{2} b_v \left(\delta_{xx} - \delta_{yy}\right) + i d_v \delta_{xy}, \\
\delta S & = -\frac{d_v}{\sqrt{2}} \left(\delta_{zx} - i \delta_{yz} \right) , \\
\delta T & = -\frac{1}{\sqrt{6}} P_{0} \sum_{j}\left(\delta_{xj} + i \delta_{yj}\right) k_j , \\
\delta U & = -\frac{1}{\sqrt{3}} P_{0} \sum_{j}\delta_{zj} k_j{.}
\end{align}
{Here,} $a_c$ and $a_v$ are the conduction- and valence-band hydrostatic deformation potentials of the host material, and $b_v$ and $d_v$ are the shear deformation potentials along the [001] and [111] directions of the host material, respectively. 
Here, we will set the deformation potentials of GaAs to be $a_c =-9.3$ eV, $a_v =-0.7$ eV, $b_v =2.0$ eV, and $d_v =5.4$ eV, following Ref.~\onlinecite{Pryor1998}.

\subsection{Light-matter interaction}
\label{sec:LightMatterInteraction}
Next, let us consider a bulk crystal of GaAs that is driven by elliptically polarized electric fields. 
Here, we take the $z$-axis ([001] direction) to be the major axis and the $x$-axis ([100] directions) to be the minor axis. 
Then, the vector potential of the elliptically polarized electric field ${\bm{A}}(t)$ can be defined as
\begin{align}
{\bm{A}}(t) &= (A_{x}(t),0,A_{z}(t)) \nonumber \\
&= A_{0} f(t) (\eta \sin{\omega t},0,\cos{\omega t}),
\end{align}
where $\eta$ and $A_{0}$ are the ellipticity and the amplitude of the electric field, respectively, and $f(t)$ is the envelop function defined as 
\begin{align}
{f(t) = \exp\left(-\frac{(t-t_0)^2}{\tau^2}\right).}
\end{align}
Here, we set $t_0=24\pi/\omega$ and $\tau=4\pi/\omega$.

We introduce the light-matter interaction through the vector potential:
\begin{align}
H_0+H_{\rm ex} &=\frac{1}{2m_0}\left(\bm{p}-\frac{e}{c} \bm{A}(t)\right)^2 \nonumber \\
&+\sum_{i}\left[ V_i(\bm{x}) + \frac{\hbar}{4m_0^2 c^2}
(\nabla V_i \times {\bm p})\cdot {\bm \sigma}
\right].
\end{align} 
Here, we have assumed that the term caused by the replacement ${\bm p} \rightarrow {\bm p}-e/c {\bm A}(t)$ in the spin-orbit interaction is small enough to be neglected. 
Thus, the Hamiltonian for the light-matter interaction is 
\begin{align}
H_{\rm ex} = - \frac{e}{m_0 c} {\bm A}(t) \cdot {\bm p}
+ \frac{e^2}{2m_0 c^2} {\bm A}^2(t).
\end{align}
The second term in $H_{\rm ex}$ can be eliminated by performing a unitary transformation $H_{\rm ex} \rightarrow U^{-1}_1 H_{\rm ex} U_1$, where 
\begin{align}
U_1 = \exp\left[\frac{ie^2}{2m_0c^2\hbar^2}\int^t_0 dt' \, {\bm A}^2(t') \right].
\end{align}
$H_{\rm ex}$ is then rewritten as $\tilde{H}_{\rm ex}$, which operates on the eigenstate $\ket{u_{n{\bm k}}}$:
\begin{align}
\tilde{H}_{\rm ex} &\equiv e^{-i {\bm{k}} \cdot {\bm{x}}} H_{\rm ex} e^{i {\bm{k}} \cdot {\bm{x}}} 
\nonumber \\
&= -\frac{e}{m_{0}c} \left[{\bm {A}}(t) \cdot \hbar \bm{k} + \bm{A}(t) \cdot {\bm{p}} \right].
\end{align}
Here, we can also eliminate the first term {in brackets} through the unitary transformation $\tilde{H}_{\rm ex} \rightarrow U_2^{-1}\tilde{H}_{\rm ex} U_2$, whose matrix elements are expressed as
\begin{align}
&\bra{u_{n{\bm k}}}U_2 \ket{u_{n'{\bm k}'}} \nonumber \\
&\hspace{5mm} =
\exp\left(-i \frac{e}{m_{0}c} \int_0^t dt' \, {\bm k}\cdot {\bm A}(t')\right) \delta_{{\bm k},{\bm k}'} \delta_{n,n'}. 
\end{align}
Thus, the Hamiltonian of the light-matter interaction becomes
\begin{align}
\tilde{H}_{\rm ex} &= -\frac{e}{m_{0}c} \bm{A}(t) \cdot {\bm{p}}.
\end{align}

\begin{figure*}[tb]
\begin{center}
\includegraphics[width=16cm]{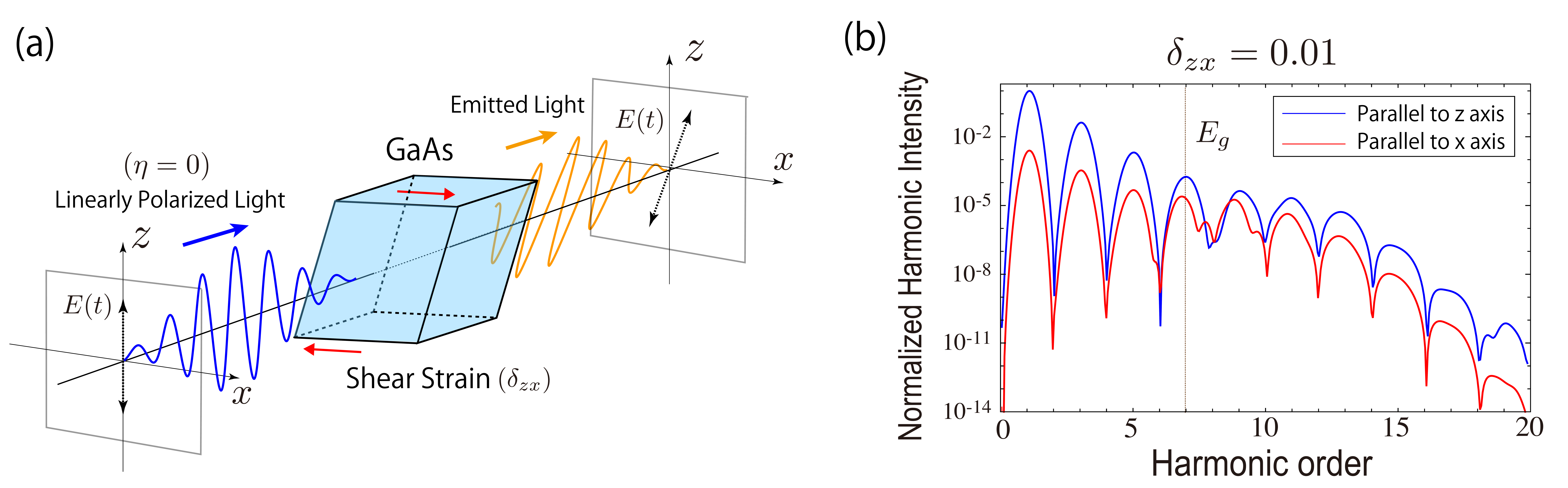}
\caption{(Color online) (a) Schematic diagram of HHG in shear-strained GaAs caused by a linearly polarized electric field ($\eta=0$). The polarization axis of the light rotates as it propagates through the shear-strained material. As a result, the electric field of the emitted light includes components along the major axis (the $z$-direction) and the minor axis (the $x$-direction). (b) Numerical result of HHG spectra {in} shear-strain{ed} GaAs caused by linearly polarized electric field ($\eta=0$) in the case of $\delta_{zx}=0.01$. The blue and red lines show the HHG spectra emitted along the major axis ($z$-axis) and minor axis ($x$-axis), respectively.
\label{fig:setup}}
\end{center}
\end{figure*}

To express the light-matter interaction in a simple matrix form, we will return to the elemental basis set of $\ket{s\!\uparrow}$, $\ket{p_x\!\uparrow}$, $\ket{p_y\!\uparrow}$, $\ket{p_z\!\uparrow}$, $\ket{s\!\downarrow}$, $\ket{p_x\!\downarrow}$, $\ket{p_y\!\downarrow}$, and $\ket{p_z\!\downarrow}$, and we will redefine them as $\ket{v_m}$ $(m=1,2,\cdots,8)$ for simplicity of notation. 
Note that the eigen wavefunctions at the $\Gamma$ point, $\ket{u_n}$'s ($n=1,2,\cdots,8)$, are expressed by a linear combination of $\ket{v_m}$ (see Appendix~\ref{app:unitary} for the explicit forms). 
In this basis set, almost all of the matrix elements in $\tilde{H}_{\rm{ex}}$ are zero because of parity symmetry. The nonzero matrix elements are given as
\begin{align}
\braket{s\sigma|\tilde{H}_{\rm{ex}}|p_x\sigma} &= ( \braket{p_x\sigma|\tilde{H}_{\rm{ex}}|s\sigma})^* \nonumber \\
& = -i \hbar \Omega_{R0} f(t) \cos \omega t, 
\label{eq:Hexelement1}
\\
\braket{s\sigma|\tilde{H}_{\rm{ex}}|p_z\sigma} &= 
(\braket{p_z\sigma|\tilde{H}_{\rm{ex}}|s\sigma})^* \nonumber \\
&=-i \hbar \Omega_{R0} \eta f(t) \sin \omega t, \label{eq:Hexelement2}
\end{align}
for $\sigma =\uparrow, \downarrow$, where we have defined the Rabi frequency as {$\Omega_{R0} = (e/c\hbar^{2})P_{0} A_{0} \equiv d_{z} E_{0}/ \hbar$}. Thus, we have derived the light-matter interaction Hamiltonian in a simple matrix form using the elemental basis set $\ket{v_m}$. Here, we set $E_{g} = 7 \hbar \omega$ and $\Omega_{R0} = 4 \omega$, respectively.
{Since the band-gap energy of GaAs is $1.42 \ {\rm{eV}}$, the frequency of the laser field is $\omega \approx 49 \ {\rm{THz}}$.
Then, the envelope parameters of the laser field{,} $t_0$ and $\tau${,} are estimated as $t_0 \approx 1.54 \ {\rm{picosecond}}$ and $\tau \approx 0.26 \ {\rm{picosecond}}$, respectively.
By assuming $d_{z} = 0.6 \ [\rm{e} \cdot \rm{nm}]$, the max intensity of the laser field {is estimated} as $E_{0} \approx 13.5 \ {\rm{MV/cm}}$.}

\subsection{Dynamical simulation}
\label{sec:dynamics}
We solve the time-dependent Schr\"odinger equation, 
\begin{align}
i \hbar \frac{\partial}{\partial t}\ket{u_{{\bm{k}}}{(t)}} &= \tilde{H}_{{\rm tot}}^{{\bm{k}}} 
\ket{u_{\bm{k}}{(t)}},
\end{align}
where $\tilde{H}_{{\rm tot}}^{{\bm{k}}} = \tilde{H}_{{\rm eff}}^{{\bm{k}}} + \tilde{H}_{{\rm ex}}$ is the total Hamiltonian. In the simulation, we employed the atomic basis $\ket{v_m}$ and expanded the wavefunction $\ket{u_{\bm{k}}{(t)}}$ as
\begin{align}
\ket{u_{\bm{k}}{(t)}} = \sum_{m=1}^8 a_{m{\bm k}}(t) \ket{v_m}.
\end{align}
Using this basis set, the matrix elements of the light-matter interaction $\tilde{H}_{{\rm ex}}$ are given by Eqs.~(\ref{eq:Hexelement1})-(\ref{eq:Hexelement2}), while those of the system Hamiltonian are given as
\begin{align}
&(\tilde{H}_{\rm eff}^{\bm k})_{mm'} \equiv
\bra{v_m} \tilde{H}_{\rm eff}^{\bm k} \ket{v_{m'}}
\nonumber \\
&\hspace{5mm} = \sum_{n,n'=1}^8 \braket{v_m | u_n}
\braket{u_n|\tilde{H}_{\rm eff}^{\bm k} |u_{n'}}
\braket{u_{n'}|v_{m'}}.
\end{align}
Here, $\braket{u_n|\tilde{H}_{\rm eff}^{\bm k} |u_{n'}}$ is as in Eq. (\ref{HH}), and $(U)_{nm}=\braket{u_{n}|v_{m}}$ is the unitary matrix for the basis transformation, whose explicit forms are in Appendix~\ref{app:unitary}. 
Thus, the time-dependent Schr\"odinger equation finally becomes
\begin{align}
i\hbar \frac{da_{m{\bm k}}}{dt} = \sum_{m'=1}^8
(\tilde{H}_{\rm eff}^{\bm k}+\tilde{H}_{\rm ex})_{mm'} a_{m'{\bm k}}(t).
\label{eq:timedepschrodinger}
\end{align}
The generated currents along the [001] and [100] directions are calculated as
\begin{align}
J_x(t) &= -{c} \Braket{\frac{\partial H_{{\rm{ex}}}}{\partial A_{x}}} \nonumber \\
&\propto -i \sum_{\bm{k} \sigma} \left[ a_{s\sigma {\bm k}}(t)^* a_{p_x\sigma {\bm k}}(t) - {\rm c.c.} \right] \\
J_z(t) &= -{c} \Braket{\frac{\partial H_{{\rm{ex}}}}{\partial A_{z}}} \nonumber \\
&\propto -i \sum_{{\bm k} \sigma} \left[ a_{s\sigma {\bm k}}(t)^* a_{p_z\sigma {\bm k}}(t) - {\rm c.c.} \right]
\end{align}
The HHG spectra in GaAs along the [001] and [100] directions {are calculated} as $I_{z}=\left|\omega {\cal J}_{z}(\omega) \right|^2$ and $I_{x}=\left|\omega {\cal J}_{x}(\omega) \right|^2$, where ${\cal J}_{z}(\omega)$ and ${\cal J}_{x}(\omega)$ are the Fourier transforms of the generated currents {$J_{z}(t)$} and {$J_{x}(t)$}.
{Here, we multiply a window function $f(t) = \exp\left(-{(t-t_0)^2}/{\tau^2}\right)$ to the generated current before its Fourier transformation.}
We numerically solve the time-dependent differential equation (\ref{eq:timedepschrodinger}) under the initial conditions where $\ket{u_{\bm k}(t=0)}=\ket{u_n}$ for occupied valence bands ($n=2,3,4,6,7,8$) and sum up the currents with respect to these six initial conditions and the Bloch wavenumber ${\bm k}$.
{We employed the fourth-order Runge-Kutta method {with a temporal mesh $\delta t = 0.05 \omega^{-1}$}.
We performed numerical integration with respect to the Bloch wavenumber using the general-purpose multidimensional integration library, CUBA\cite{CUBA,HAHN200578}.}

\section{RESULTS AND DISCUSSION}
\label{RESULTS AND DISCUSSIONS}
In this section, we discuss the characteristics of HHG originating from shear-strained GaAs. 
First, let us examine the numerical results of HHG caused by a linearly polarized electric field in Sec.~\ref{Linearly polarized electric fields}.
Here, we identify a rotation of the polarization axis of the emitted light. 
Next, let us examine the numerical results of HHG caused by elliptically polarized light in Sec.~\ref{Elliptically polarized electric fields}.
These results indicate a breakdown in reciprocity of HHG in the shear-strained material. 
Sec.~\ref{Breakdown of the reciprocity} discusses the physical interpretation of the numerical results.
{In the appendix, we provide information of the ellipticity of generated high harmonics.}

\subsection{Linearly polarized electric fields}
\label{Linearly polarized electric fields}
Now, let us consider the case of linearly polarized electric fields ($\eta = 0$) in shear-strained GaAs:
\begin{align}
\delta_{ij} = \left\{ \begin{array}{ll}
\delta_{zx} \ne 0, & (i,j)=(z,x), \\
0, & ({\rm otherwise}). \end{array} \right.
\end{align}
Fig.~\ref{fig:setup}~(a) shows a schematic diagram of the effect of strain on HHG ($\delta_{zx}\ne 0$). 
When the electric {field} of the incident light {is} polarized in the $z$-direction, current is generated along the major ($z$) and minor ($x$) axes (see also Fig.~\ref{fig:config1}). 
As a result, the electric fields of the emitted light also include an $x$-component, resulting in a rotation of the polarized light.

\begin{figure}[tb]
\begin{center}
\includegraphics[width=8cm]{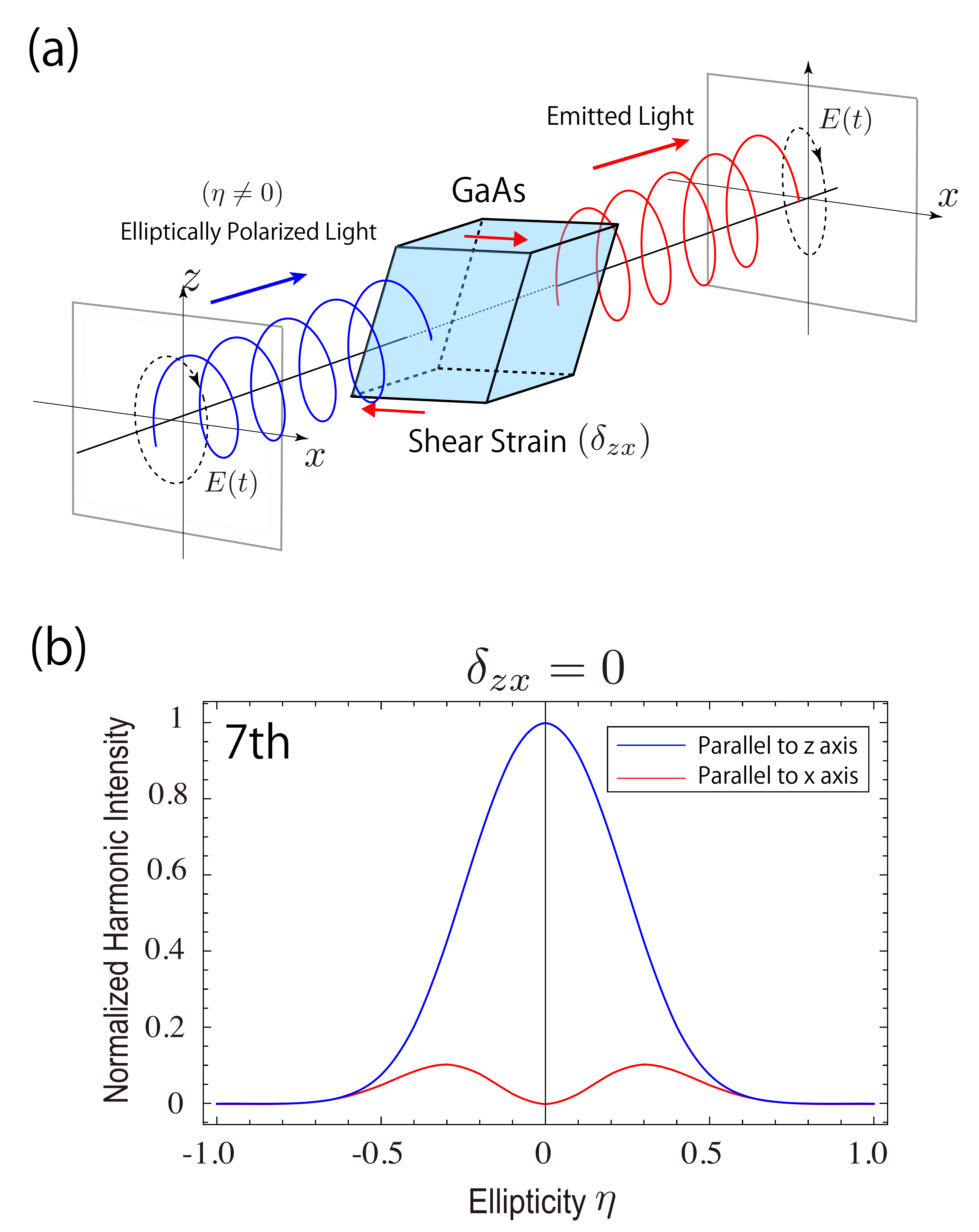}
\caption{(Color online) (a) Schematic diagram of HHG {for} shear-strain{ed} GaAs caused by elliptically polarized electric fields ($\eta \neq 0$). (b) {Numerical result for t}he {calculated} ellipticity dependence of HHG {in the absence of strain}. The blue and red lines show the ellipticity dependences of HHG emitted along to the major axis ($z$-axis) and minor axis ($x$-axis), respectively. The finite shear strain changes {these} dependences and is expected to {cause optical activity, i.e., different responses to} the right-handed ($\eta>0$) and left-handed ($\eta<0$) elliptically polarized electric fields.
\label{fig:elliptic}}
\end{center}
\end{figure}

\begin{figure*}[!p]
\begin{center}
\includegraphics[width=16cm]{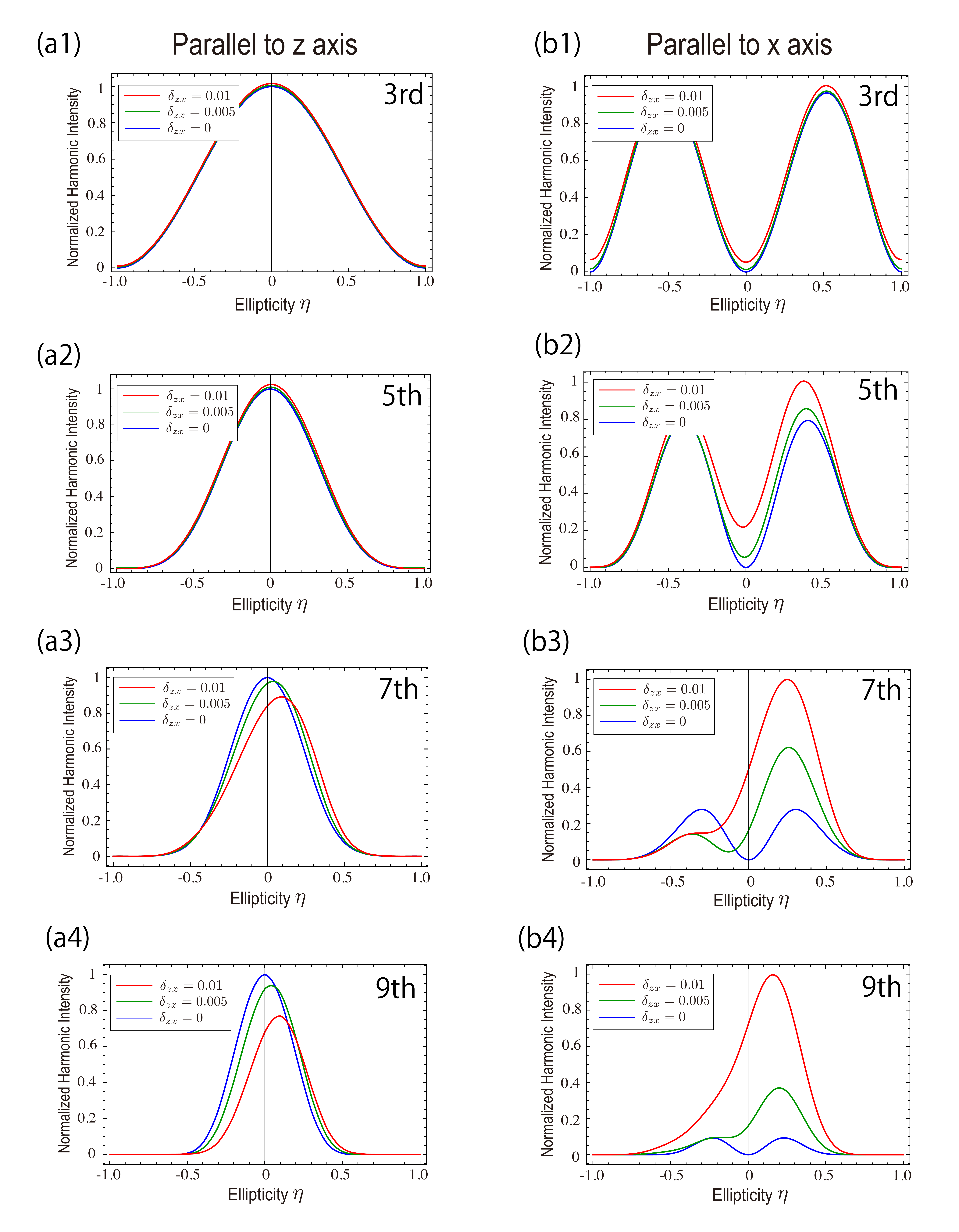}
\caption{(Color online) Calculated ellipticity dependences of high-order harmonic intensities focusing on the third ((a1) and (b1)), fifth ((a2) and (b2)), seventh ((a3) and (b3)), and ninth harmonics ((a4) and (b4)). Figures (a1)-(a4) show the harmonic intensities emitted along the major axis ($z$-axis), while figures (b1)-(b4) show those along the minor axis ($x$-axis). The red, green, and blue lines indicate the {ellipticity dependences} for $\delta_{zx}{=0}${, 0.005, and 0.01, respectively}. These figures indicate that the right-handed ($\eta >0$) and left-handed ($\eta <0$) elliptically polarized electric fields yield the different HHG spectra. This breakdown of reciprocity becomes significant for the high-order harmonics above the band-gap energy (n $\geq$ $E_g/\hbar {\omega} =7$).
\label{fig:elliptic2}}
\end{center}
\end{figure*}

Fig.~\ref{fig:setup}~(b) shows the calculated HHG spectra for shear-strained GaAs ($\delta_{zx}=0.01$). 
The blue and red curves denote the intensity of the emitted light parallel to the major ($z$) and minor ($x$) axes, respectively.
{We note that a few megapascal pressure yields one percent displacement of the lattice vectors in GaAs ($\delta_{zx}=0.01$), that can be realized in  an experiment by using the boat technique or the liquid encapsulated Czochralski technique\cite{yonenaga1992impurity}.}
The spectra for the $z$-direction (the major axis) exhibit several features characteristics of HHG; the peaks correspond to $n\omega$ for odd $n$, and their heights first decay exponentially with respect to $n$ in the perturbative regime ($n<7$), eventually reaching a plateau in the intermediate regime ($7 \le n \le 13$), and finally collapsing exponentially again for large $n$ ($>13$). 
The spectra for the $x$-direction (the minor axis) are similar to those in the $z$-direction except that their intensity is much smaller. 
Here, it is remarkable that the ratio of the intensity in the $x$-direction relative to that of the $z$-direction is enhanced around the band-gap energy, that is, $n\omega \approx E_{\rm g}$, where $E_{\rm g}$ is {the} band gap energy. 
To discuss this strain-induced effect in detail, the next subsection examines the case of elliptically polarized electric fields.

\subsection{Elliptically polarized electric fields}
\label{Elliptically polarized electric fields}
Next, let us discuss the case of elliptically polarized electric fields ($\eta \ne 0$). Figure~\ref{fig:elliptic}~(a) shows a schematic diagram of the strain-induced effect on GaAs driven by elliptically polarized electric fields ($\delta_{zx} \ne 0$). 
The $z$- and $x$-components of the electric fields of the emitted light are modified from those of the incident light. 
The intensities of the emitted light in the two directions are plotted as a function of ellipticity $\eta$. As a reference, Fig.~\ref{fig:elliptic}~(b) plots the ellipticity dependences of the seventh harmonics emitted along to the $z$-axis (blue line) and $x$-axis (red line) for unstrained GaAs ($\delta_{zx}=0$). 
In this figure, we can identify that the HHG intensity in the $z$-direction (the major axis) has a single Gaussian-like peak at $\eta = 0$, while the intensity in the $x$-direction (the minor axis) has two peaks at finite values of $\eta$. 
These features have already been identified in the previous work \cite{Tamaya2016PRBR}; the double peaks in the $x$-direction grow with increasing field intensity and become especially pronounced in the semimetal regime\cite{Tamaya2016}. 
{T}his phenomenon was observed in an experiment on HHG using graphene and $\rm{MoS_2}$\cite{Yoshikawa2017}.

Fig.~\ref{fig:elliptic2} shows high-order harmonic intensities of the third ((a1) and (b1)), fifth ((a2) and (b2)), seventh ((a3) and (b3)), and ninth harmonics ((a4) and (b4)) as a function of ellipticity $\eta$ for {three} values of $\delta_{zx}$. 
The red, green, and blue lines indicate the HHG spectra for $\delta_{zx}=0.01$, $0.005$, and $0$, respectively. 
For unstrained GaAs ($\delta_{zx}=0$), the HHG intensity is symmetric with respect to an inversion of ellipticity ($\eta \rightarrow -\eta$). 
This inversion symmetry, however, is broken for strained GaAs ($\delta_{zx} = 0.005,0.01$). 
The peak position in the $z$-direction (the major axis) shifts toward positive $\eta$ (see figures (a1)-(a4)). 
As the order of HHG increases, the peak shift becomes more significant, and its height gradually decreases in comparison with the unstrained case.
For the $x$-direction (the minor axis), the heights of the two peaks become different in the strained case (see figures (b1)-(b4)). 
From the numerical results of the seventh and ninth HHG (see figures (b3) and (b4)), we find that one of the two peaks disappears for sufficiently large strain.
{We note that the ellipticity dependences of HHG for $\delta_{zx}=-\delta$ coincides with the result for $\delta_{zx}=\delta$ by reversing the ellipticity ($\eta \rightarrow -\eta$).}

{
It is known that a circularly polarized electric field have a strict selection rule that completely suppresses the all-order harmonics in atomic system.
In our numerical calculation {for} GaAs, we could identify {{almost} complete suppression of high harmonics for the unstained case.
For the strained case, however, this selection rule is broken down.
This can be identified in figure (b1); a finite intensity of HHG exists even for a circularly polarized field ($\eta = \pm 1$) when $\delta_{zx} \ne 0$.
This breakdown of the selection rule originates from the symmetry reduction of the crystal structure, that would be discussed in the next subsection.
}}

\subsection{Breakdown of reciprocity}
\label{Breakdown of the reciprocity}
The breakdown of the reciprocity relation between the right-handed ($\eta >0$) and left-handed ($\eta <0$) elliptically polarized electric fields obtained in this work can be understood in terms of symmetry reduction of the crystal structure. 
The external shear strain changes the crystal structure of GaAs from cubic (zinc blende structure) into monoclinic, inducing off-diagonal elements in the dielectric tensor $\epsilon_{zx}$\cite{Boyd2008,powell2010symmetry}.
The emergence of $\epsilon_{zx}$ directly means optical activity, i.e., different responses to left-handed and right-handed elliptically polarized electric fields. It should be noted that the breakdown in reciprocity becomes rather pronounced when the emitted photon energy exceeds the band-gap energy, which is $E_{g} = 7\hbar {\omega}$ in {our} calculation. 
The origin of this feature is conjectured to be as follows. 
When the emitted photon energy is larger than the band-gap energy, the number of excitation channels relevant to HHG largely increases in comparison with those for the low-order harmonics below the band-gap energy. 
The increase in the available channels contributes to emergence of the plateau structure in the HHG spectra\cite{Tamaya2019}, as well as oscillatory behavior as a function of field strength\cite{xia2020highharmonic}.
Thus, we suppose that the sensitivity of the high-order harmonics above the band-gap energy to the external strain effect is caused by an increase in the excitation channels.
This breakdown of reciprocity in HHG that is sensitive to shear strain may be used for applications such as spatially-resolved measurement of lattice deformation. 
The features revealed here may also be useful for developing a mechanical control of HHG ellipticity.
{
{This sensitivity can be controlled by tuning the ratio $E_{g}/{\hbar \omega}$; the emitted lower-order harmonics near the band-gap energy are expected to show the sensitivity for the shear-strain effect.}
{For a short pulse, another possibility for controlling the piezo-optic effect is carrier envelope phase (CEP), which is the offset angle between the field envelope and the carrier wave.}
{Although the CEP effect may be important for application of the piezo-optic effect, it is beyond the scope of this paper.}
{We would like to leave it as a future problem.}}

Note that the strain effect (piezo-optic effect) should appear in other semiconductors, such as AlAs and InAs, in accordance with the same formulation based on the Luttinger-Kohn-Bir-Pikus model. 
By performing numerical calculations with different Luttinger parameters, we can easily identify similar properties of HHG in these materials. 
We also suppose that the piezo-optic effect obtained here is not specific to materials covered by the Luttinger-Kohn-Bir-Pikus model and that it appears in various systems, because the symmetry reduction of the lattice structure can be induced by shear strain. 
For a general discussion, we need to extend our theory to take the details of the band structure of the materials as well as the corresponding Bloch wavefunctions into account. This will be left as a future problem.

{W}e {also} comment on how the non-perturbative effect in HHG is important for obtaining a large piezo-optical effect. 
According to the previous work\cite{Tamaya2016PRBR,Yoshikawa2017}, the ellipticity dependence of HHG parallel to the minor axis ($x$-axis) is not so pronounced in the multiphoton absorption {(perturbative)} regime.
Therefore, to obtain a sufficiently large piezo-optical effect, we have to apply a strong field so that non-perturbative regimes, which were identified as the AC Zener {or} the semimetal regimes in Ref.~\onlinecite{Tamaya2016}, are realized.
{{For} GaAs, this threshold intensity between {the} perturbative and non-perturbative regime{s} could be estimated at around several MV/cm\cite{xia2020highharmonic}.}
Because the amplitude of the perturbation is roughly estimated by the ratio between the field intensity {(the Rabi frequency)} and the band-gap energy, it is {expected} that the piezo-optical effect discussed here can be observed more clearly by using narrow-gap and zero-gap semiconductors, such as InSb and graphene.

{Finally, we will refer to influence of the inter-particle Coulomb interactions on HHG in GaAs.
Major effects of the inter-particle Coulomb interaction are a band-gap renormalization and scattering between excited electrons (holes).
Both effects are expected to be weak in GaAs, because the dielectric constant is large ($\epsilon_r \simeq 13$), indicating large screening effect which weakens inter-particle Coulomb interaction.
{T}he former effect is regarded as a variation in the band-gap energy, whose order is, at most,  several meV\cite{PhysRevB.69.205204}, while the latter can be taken into account in terms of relaxation/dephaing effect that {does not} influence so much on the nonlinear optical processes\cite{Shen1984,Yariv1984,Boyd2008}.
{In addition, since the Coulomb interaction equally affects each harmonic, the results of this paper would not be changed qualitatively.}
}

\section{Conclusion}
\label{Conclusion}
We theoretically investigated the shear-strain effect of HHG in GaAs. 
By constructing a theoretical framework based on the Luttinger-Kohn-Bir-Pikus model, we calculated the spectra of HHG emissions parallel to the major and minor axes. 
Our numerical results for linearly polarized incident light implied that shear-stained materials have optical activity (non-reciprocity), i.e., different responses to right-handed and left-handed elliptically polarized electric fields. 
To verify this conjecture, we calculated the ellipticity dependence of HHG emitted from shear-strained GaAs. 
Consequently, we found a breakdown in reciprocity with respect to inversion of the ellipticity. 
We also found that this breakdown is much more pronounced for higher-order harmonics that exceed the band-gap energy. 
These features can be understood in terms of a reduction in the symmetry of the crystal structure of GaAs inducing off-diagonal elements in the dielectric tensor $\epsilon_{zx}$. 
The conclusions presented in this paper are generally applicable to various semiconductors because the symmetry reduction can be induced by the shear-strain effect. 
Our study thus provides a foundation for strain engineering of nonlinear optics, including HHG, that is impossible in gaseous media.

\vspace{2mm}
T. T. and T. K. gratefully acknowledge support from the Japan Society for the Promotion of Science (JSPS KAKENHI Grants No. JP19K14624 and No. JP20K03831). 

\appendix

\section{Basis transformation}
\label{app:unitary}
The eigen wavefunctions at the $\Gamma$ point, $\ket{\Psi_{n{\bm k}=0} }= \ket{u_n}$ ($n=1,2,\cdots,8$) are given as a superposition of atomic orbitals. In the presence of the spin-orbit interaction, they are categorized by $\ket{J,{J}_{z}}$, where $J$ is the total orbital angular momentum and $J_z$ is the $z$-component of the angular momentum:
\begin{align}
\ket{u_{1}}& \equiv \Ket{\frac{1}{2},+\frac{1}{2}} = \ket{s\! \uparrow}, \nonumber \\
\ket{u_{2}}& \equiv \Ket{\frac{3}{2},+\frac{3}{2}} = \frac{i}{\sqrt{2}} (\ket{p_x\! \uparrow} + i\ket{p_y\! \uparrow}), \nonumber \\
\ket{u_{3}}& \equiv \Ket{\frac{3}{2},+\frac{1}{2}} = \frac{i}{\sqrt{6}} (\ket{p_x\! \downarrow} + i\ket{p_y\! \downarrow} -2 \ket{p_z\! \uparrow}), \nonumber \\
\ket{u_{4}}& \equiv \Ket{\frac{1}{2},+\frac{1}{2}} = \frac{i}{\sqrt{3}} (\ket{p_x\! \downarrow} + i\ket{p_y\! \downarrow} + \ket{p_z\! \uparrow}). \nonumber
\end{align}
Here, we define $\ket{s\sigma}$, $\ket{p_x\sigma}$, $\ket{p_y\sigma}$, and $\ket{p_z\sigma}$ to be the $s$-, $p_{x}$-, $p_{y}$-, and $p_{z}$-like wavefunctions, respectively, for spin components $\sigma = \uparrow$ or $\downarrow$. The remaining set of Bloch basis states are expressed as
\begin{align}
\ket{u_{5}}& \equiv \Ket{\frac{1}{2},-\frac{1}{2}} = -\ket{s \! \downarrow}, \\
\ket{u_{6}}& \equiv \Ket{\frac{3}{2},-\frac{3}{2}} = -\frac{i}{\sqrt{2}} (\ket{p_x\! \downarrow} - i\ket{p_y\! \downarrow}), \\
\ket{u_{7}}& \equiv \Ket{\frac{3}{2},-\frac{1}{2}} = \frac{i}{\sqrt{6}} (\ket{p_x\! \uparrow} - i\ket{p_y \! \uparrow} +2 \ket{p_z\! \downarrow}), \\
\ket{u_{8}}& \equiv \Ket{\frac{1}{2},-\frac{1}{2}} = \frac{i}{\sqrt{3}} (\ket{p_x\! \uparrow} - i\ket{p_y\! \uparrow} - \ket{p_z\!\downarrow}). 
\end{align}
{From these expressions, one can easily obtain the matrix element of the unitary operation, $(U)_{nm}=\braket{u_{n}|v_{m}}$.}

{\section{Ellipticity of emitted harmonics}}

\begin{figure}[tb]
\begin{center}
\includegraphics[width=8cm]{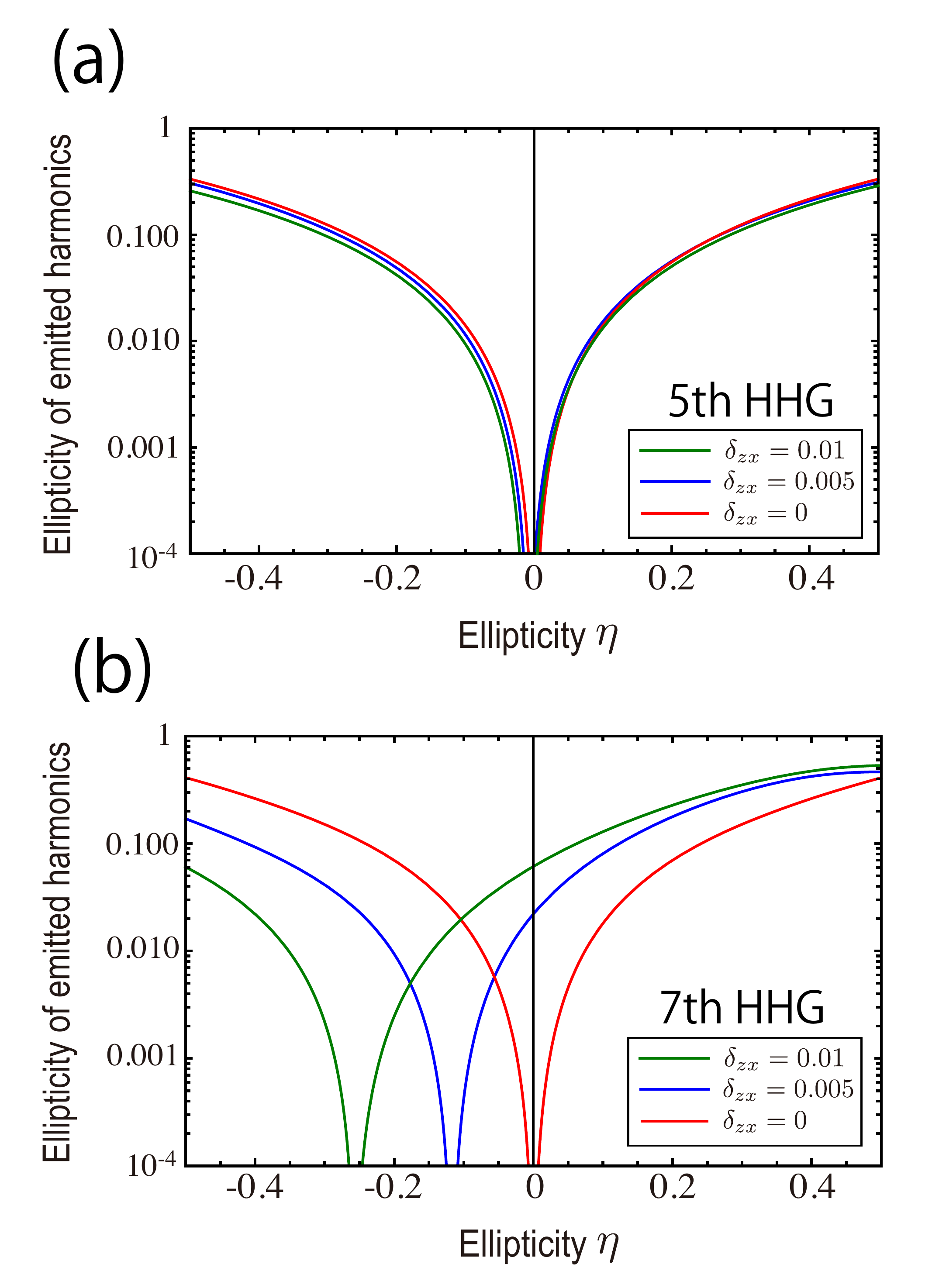}
\caption{(Color online) {Ellipticity of emitted fifth- (a) and seventh-order harmonics (b) as a function of the ellipticity of the incident electric field. 
The red, blue and green curves denote the ellipticity of the harmonics for the cases of $\delta_{zx}=0$, $0.005$, and $0.01$, respectively.} 
\label{fig:elliptic3}}
\end{center}
\end{figure}

{In this {appendix}, we discuss the ellipticity of emitted harmonics.
{The ellipticity of the emitted $n$th-order harmonics is defined as $\epsilon=|A/B|$, where $A$ and $B$ are the amplitude of the electric field of the semi-major and semi-minor axes for the elliptic light:}
\begin{align}
A &= |\bm{J}^{n\rm{th}}(\omega)|\sqrt{\frac{1+\sqrt{1-\sin^2{(2\theta)}\sin^2{\beta}}}{2}}, \nonumber \\
B &= |\bm{J}^{n\rm{th}}(\omega)|\sqrt{\frac{1-\sqrt{1-\sin^2{(2\theta)}\sin^2{\beta}}}{2}}. \nonumber 
\end{align}
{
Here, $\bm{J}^{n\rm{th}}(\omega) = \left(J^{n\rm{th}}_{z}(\omega),J^{n\rm{th}}_{x}(\omega)\right) = \left(\left|\bm{J}^{n\rm{th}}(\omega)\right|\cos\theta, \left|\bm{J}^{n\rm{th}}(\omega)\right| \sin \theta\right)$ is the Fourier transforms of the generated currents $J^{n\rm{th}}_{z}(t)$ and $J^{n\rm{th}}_{x}(t)$, and $\beta$ is a difference between arguments of $J^{n\rm{th}}_{z}(\omega)$ and $J^{n\rm{th}}_{x}(\omega)$, that is, $\beta = {\rm{Arg}}\left[J^{n\rm{th}}_{z}(\omega)\right]-{\rm{Arg}}\left[J^{n\rm{th}}_{x}(\omega)\right]$.

}

We show in Fig.~\ref{fig:elliptic3}~(a) and (b) the ellipticity of emitted fifth and seventh-order harmonics for the cases of $\delta_{zx}=0$ (red line), $0.005$ (blue line) and $0.01$ (green line), respectively.
{These figures indicate that the ellipticity of the emitted harmonics vanishes at $\eta=0$ for the unstrained case (red lines).}
{For the strained cases (blue and green lines), the value of $\eta$ at which the ellipticity of the emitted harmonics vanishes shifts toward the negative direction.
This shift is emphasized for the seventh-order harmonics in comparison with the fifth-order one.}
{These features may be helpful for a control of the emitted HHG in the shear-strain semiconductors.}
}

\bibliography{GaAsHHGref.bib}
\end{document}